%% file: main.tex
\newif\ifshort\shorttrue

\documentclass[sigplan]{acmart}
\acmConference[PL]{~}{~}{~}
\acmYear{2018}
\acmISBN{} % \acmISBN{978-x-xxxx-xxxx-x/YY/MM}
\acmDOI{} % \acmDOI{10.1145/nnnnnnn.nnnnnnn}
\startPage{1}

\setcopyright{none}
\setcitestyle{nosort}      %% With 'acmnumeric', to disable automatic
\usepackage{ifthen}
\usepackage{stmaryrd}
\usepackage{xspace}
\usepackage{tikz}
\usetikzlibrary{calc,positioning}
\usepackage{paralist}

\graphicspath{{./figures/}}

\input{macros}

\begin{document}

%% Title information
\title[SAS Games]{Combinations of Qualitative Winning for \\ Stochastic Parity Games}         %% [Short Title] is optional;
                                        %% when present, will be used in
                                        %% header instead of Full Title.
%\titlenote{with title note}             %% \titlenote is optional;
                                        %% can be repeated if necessary;
                                        %% contents suppressed with 'anonymous'
%\subtitle{Subtitle}                     %% \subtitle is optional
%\subtitlenote{with subtitle note}       %% \subtitlenote is optional;
                                        %% can be repeated if necessary;
                                        %% contents suppressed with 'anonymous'

%% Author information
%% Contents and number of authors suppressed with 'anonymous'.
%% Each author should be introduced by \author, followed by
%% \authornote (optional), \orcid (optional), \affiliation, and
%% \email.
%% An author may have multiple affiliations and/or emails; repeat the
%% appropriate command.
%% Many elements are not rendered, but should be provided for metadata
%% extraction tools.

%% Author with single affiliation.
\author{Krishnendu Chatterjee}
%\authornote{with author1 note}          %% \authornote is optional;
                                        %% can be repeated if necessary
%\orcid{nnnn-nnnn-nnnn-nnnn}             %% \orcid is optional
\affiliation{
%  \position{Position1}
%  \department{Department1}              %% \department is recommended
  \institution{IST Austria}            %% \institution is required
%  \streetaddress{Street1 Address1}
  \city{Klosterneuburg}
%  \state{State1}
%  \postcode{Post-Code1}
  \country{Austria}                    %% \country is recommended
}
\email{krish.chat@ist.at}          %% \email is recommended

%% Author with two affiliations and emails.
\author{Nir Piterman}
%\authornote{with author2 note}          %% \authornote is optional;
                                        %% can be repeated if necessary
%\orcid{nnnn-nnnn-nnnn-nnnn}             %% \orcid is optional
\affiliation{
%  \position{Position2a}
%  \department{Department2a}             %% \department is recommended
  \institution{University of Leicester}           %% \institution is required
%  \streetaddress{Street2a Address2a}
  \city{Leicester}
%  \state{State2a}
%  \postcode{Post-Code2a}
  \country{UK}                   %% \country is recommended
}
\email{nir.piterman@gmail.com}         %% \email is recommended

%% Abstract
%% Note: \begin{abstract}...\end{abstract} environment must come
%% before \maketitle command
\begin{abstract}
\input{abstract}

\end{abstract}

%% 2012 ACM Computing Classification System (CSS) concepts
%% Generate at 'http://dl.acm.org/ccs/ccs.cfm'.

%% End of generated code

%% Keywords
%% comma separated list
  %% \keywords are mandatory in final camera-ready submission

%% \maketitle
%% Note: \maketitle command must come after title commands, author
%% commands, abstract environment, Computing Classification System
%% environment and commands, and keywords command.
\maketitle

\newcommand\inc\input
\inc{intro}
\inc{background}
\inc{sasmdp}
\inc{sas}
\inc{sls}
\inc{conc}

%% Acknowledgments
%\begin{acks}                            %% acks environment is optional
%                                        %% contents suppressed with 'anonymous'
%  %% Commands \grantsponsor{<sponsorID>}{<name>}{<url>} and
%  %% \grantnum[<url>]{<sponsorID>}{<number>} should be used to
%  %% acknowledge financial support and will be used by metadata
%  %% extraction tools.
%  This material is based upon work supported by the
%  \grantsponsor{GS100000001}{National Science
%    Foundation}{http://dx.doi.org/10.13039/100000001} under Grant
%  No.~\grantnum{GS100000001}{nnnnnnn} and Grant
%  No.~\grantnum{GS100000001}{mmmmmmm}.  Any opinions, findings, and
%  conclusions or recommendations expressed in this material are those
%  of the author and do not necessarily reflect the views of the
%  National Science Foundation.
%\end{acks}

%% Bibliography
%%\bibliographystyle{abbrv}
%\bibliography{ok}

\clearpage
\onecolumn
% Appendix
\appendix

\input{appendix}

\end{document}

%% file: macros.tex
\newcommand{\pzero}{{Player~0}\xspace}
\newcommand{\pone}{{Player~1}\xspace}

\ifshort
\newcommand{\shorten}[2]{#2}
\newcommand{\shortv}[1]{#1}
\else
\newcommand{\shorten}[2]{#1}
\newcommand{\shortv}[1]{}
\fi

\usetikzlibrary{shapes}
\newcommand{\tikzpathdefaults}{\path[every node/.style={font=\sffamily\small}]}
\newcommand{\gamedefaults}[2]{
\begin{center}
  \begin{tikzpicture}[shorten >=1pt,node distance=#1mm, inner sep=1pt,
      minimum size=6mm,auto, thick, ->]
#2
\end{tikzpicture}
\end{center}
}

\newcommand{\game}[1]{\gamedefaults{15}{#1}}

\newcommand{\circlenode}[4]{
\ifthenelse{\equal{#2}{}}
{\node (#1) [draw,circle] {$#3$};}
{\node (#1) [draw,circle,#2] {$#3$};}
\ifthenelse{\equal{#4}{}}
{}
{\node (label#1) [right=1pt of #1,yshift=1pt] {$#4$};}
}

\newcommand{\rectanglenode}[4]{
\ifthenelse{\equal{#2}{}}
{\node (#1) [draw,rectangle] {$#3$};}
{\node (#1) [draw,rectangle,#2] {$#3$};}
\ifthenelse{\equal{#4}{}}
{}
{\node (label#1) [right=2mm of #1,yshift=2mm] {$#4$};}
}

\newcommand{\diamondnode}[4]{
\ifthenelse{\equal{#2}{}}
           {\node (#1) [draw,diamond] {$#3$};}
           {\node (#1) [draw,diamond,#2] {$#3$};}
           \ifthenelse{\equal{#4}{}}
                      {}
                      {\node (label#2) [right=2mm of #1,yshift=2mm]
                        {$#4$};}
                      }

\newcommand{\labeln}[3]{
{\node (label#1) [#2] {#3};}
}

\newcommand{\pznode}[4]{\circlenode{#1}{#2}{#3}{#4}}
\newcommand{\ponode}[4]{\rectanglenode{#1}{#2}{#3}{#4}}

\newcommand{\pz}[3]{\pznode{#1}{#2}{#3}{}}
\newcommand{\po}[3]{\ponode{#1}{#2}{#3}{}}

\newcommand{\fulledge}[5]{
\tikzpathdefaults
(#1) edge [#4] node [#5] {#3} (#2);
}

\newcommand{\edge}[2]{
\fulledge{#1}{#2}{}{}{}
}

\newcommand{\edgestosucc}[1]{
{
\node (ls#1) [below of=#1,yshift=2mm] {$_{succ(v)}$};
\node (lsr#1) [right=0.1mm of ls#1,xshift=-1mm,yshift=-2mm] {};
\node (lsl#1) [left=0.1mm of ls#1,xshift=1mm,yshift=-2mm] {};
\edge{#1}{lsr#1}
\edge{#1}{lsl#1}
}
}

\renewcommand{\comment}[1]{}

%% file: abstract.tex
We study Markov decision processes and turn-based stochastic games 
with parity conditions. 
There are three qualitative winning criteria, namely, sure winning,
which requires all paths must satisfy the condition, almost-sure winning,
which requires the condition is satisfied with probability~1, and limit-sure
winning, which requires the condition is satisfied with probability arbitrarily
close to~1.
We study the combination of these criteria for parity conditions, e.g., 
there are two parity conditions one of which must be won surely, and the other
almost-surely.
The problem has been studied recently by Berthon et.~al for MDPs with 
combination of sure and almost-sure winning, under infinite-memory strategies, 
and the problem has been established to be in NP $\cap$ coNP.
Even in MDPs there is a difference between finite-memory and infinite-memory 
strategies.
Our main results for combination of sure and almost-sure winning are as follows: 
(a)~we show that for MDPs with finite-memory strategies the problem lie in NP $\cap$ coNP;
(b)~we show that for turn-based stochastic games the problem is coNP-complete, both for 
finite-memory and infinite-memory strategies;
and (c)~we present algorithmic results for the finite-memory case, both for MDPs and 
turn-based stochastic games, by reduction to non-stochastic parity games.
In addition we show that all the above results also carry over to combination of sure
and limit-sure winning, and results for all other combinations can be derived from 
existing results in the literature.
Thus we present a complete picture for the study of combinations of qualitative 
winning criteria for parity conditions in MDPs and turn-based stochastic games.

%Recently Berthon et.~al considered the problem of enforcing sure
%winning one parity conditions while simulataneously enfocing
%almost-sure winning according to another parity condition.
%They have considered the case of MDPs and showed that ...
%Here, we revisit this problem for MDPs and for games.
%We show that already for MDPs there is a big difference between the
%cases where we allow infinite-memory vs.~finite-memory.
%We also show that the case of allowing only finite-memory can be
%solved effectively by considering a reduction to parity games and that
%these techniques extend beyond MDPs to two-player stochastic games.

%% file: appendix.tex
\section{Appendix}

\subsection{Proofs from Section~\ref{section:sas MDPs}}

\shortv{
  \noindent
  We include the proof of Theorem~\ref{theorem:sas mdp optimal not
    finite}:

  \proofoftheoremsasmdpoptimalnotfinite
}

\shortv{
  \noindent
  We include the proof of Lemma~\ref{lemma:sas mdp to buchi sas mdp}:

  \proofoflemmasasmdptobuchisasmdp
}

\noindent
We include the proof of Lemma~\ref{lemma:buchi sas mdp to parity and
  buchi}:

\begin{proof}
  Consider an MDP $G=(V,(V_0,V_p),E,\kappa,\mathcal{W})$, where
  $\mathcal{W}=(W_s,W_{as})$ is such that $W_{as}$ is a B\"uchi
  condition.
  Let $B\subseteq V$ denote the set of B\"uchi configurations of the
  almost-sure winning condition.
  Let $p_s:V\rightarrow [1..d_s]$ be the parity ranking function that
  induces $W_s$.
  We create the (non-stochastic) game $G'$ by replacing every configuration
  $v\in V_p$ by one of the gadgets in Figure~\ref{figure:reduction
    from almost sure to sure mdp}.
  Formally, $G'=(V',(V'_0,V'_1),E,\mathcal{W'})$, where the components of
  $G'$ are as follows:
  \begin{itemize}
  \item
    $V'_0 = V_0 \cup \{ (\tilde{v},0), (\tilde{v},2),(\hat{v},1) ~|~
    v\in V_p\}$
  \item
    $V'_1 = \{\overline{v}, (\hat{v},0) ~|~ v\in V_p\}$
  \item
    $E'=\{ (v,w) ~|~ (v,w)\in E \cap (V_0\times V_0)\}$ \shorten{\hfill}{} $\cup\hspace{1pt}$

    $
    \begin{array}{l l}
    \{ (v,\overline{w}) ~|~ (v,w)\in E \cap (V_0 \times V_p) \} & \cup \\
    \{ ((\hat{v},i),w) ~|~ (v,w)\in E \cap (V_p \times V_0)\} & \cup \\
    \{ ((\hat{v},i),\overline{w}) ~|~ (v,w)\in E \cap (V_p \times V_p)\} & \cup \\
    \{ (\overline{v},(\tilde{v},0)),%(\overline{v},(\tilde{v},2)),
        ((\tilde{v},0),(\hat{v},0)), ((\tilde{v},2),(\hat{v},1)) ~|~
        v\in V_p \}  & \cup \\
      \{ (\overline{v},(\tilde{v},2)) ~|~ v\in
      V_p\cap B\}
    \end{array}
    $
  \item
    $\mathcal{W'}$ is the winning condition resulting from the
    conjunction (intersection) of the B\"uchi condition $V_0\cap B
    \cup \{(\hat{v},0) ~|~ v\in V_p\}$ and the parity condition
    induced from $\alpha':V'\rightarrow [1..d_s]$, where
    $\alpha'(v)=\alpha_{s}(v)$ for $v\in V_0$ and
    $\alpha'(\overline{v})=\alpha'(\tilde{v},i)=\alpha'(\hat{v},i)=\alpha_s(v)$.
  \end{itemize}

  \begin{figure}[bt]
    \input{figures/gadgetmdp.tex}
    \vspace*{-7mm}
    \caption{\label{figure:reduction from almost sure to sure mdp}%
      Gadget to replace probabilistic configurations in the reduction.}
  \vspace*{-4mm}
  \end{figure}
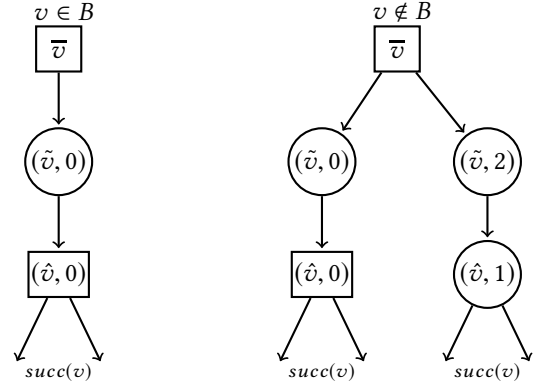

  We show that if \pzero has a finite-memory winning strategy in $M$
  she wins in $G$.
  Consider the combination of the winning strategy for \pzero in $M$
  with $M$.
  It follows that every bottom SCC is winning according to both $W_s$
  and $W_{as}$.
  We define the following strategy in $G'$ that uses the same memory.
  We have to extend the strategy by a decision from configurations of the form
  $(\hat{v},1)$, which belong to \pzero in $G'$.
  If the configuration $(\hat{v},1)$ is in a bottom SCC, \pzero chooses the
  successor of $v$ that minimizes the distance to $B$.
  By assumption every bottom SCC has some configuration in $B$
  appearing in it.
  If the configuration $(\hat{v},1)$ is in a non-bottom SCC, \pzero chooses the
  successor of $v$ with minimal distance to exit the SCC.
  Consider a play consistent with this strategy.
  Clearly, this play is possible also in the MDP $G$. By assumption
  the play satisfies $\alpha_s$.
  If a play visits infinitely often configurations of the type $(\hat{v},0)$,
  then this play satisfies the B\"uchi winning condition.
  Consider a play that visits finitely often configurations of the type
  $(\hat{v},0)$.
  It follows that whenever it visits a configuration of the type $\overline{v}$
  the distance to leave non-bottom SCC decreases and hence the play
  eventually reaches a bottom SCC.
  Similarly, after reaching a bottom SCC whenever a configuration of the type
  $\overline{v}$ is visited the distance to $B$ decreases.

  We show that if \pzero wins the game $G'$ she has a finite-memory
  winning strategy in $M$.
  As $G'$ does not have stochastic elements and as the winning
  condition there is $\omega$-regular, \pzero has a finite-memory pure
  winning strategy.
  In case that from $\overline{v}$ \pone chooses the successor
  $(\tilde{v},0)$ then the strategy instructs \pzero what to do from
  every successor of $v$.
  In case that from $\overline{v}$ \pone chooses the successor
  $(\tilde{v},2)$ then the strategy instructs \pzero to choose a
  unique successor of $v$.
  The strategy of \pzero in $M$ follows the strategy of \pzero in $G$.
  Whenever after a configuration $v\in V_p$ we discover that the
  successor is the one that is consistent with the choice from
  $(\hat{v},1)$ we update the memory accordingly.
  Whenever after a configuration $v\in V_p$ we discover that the
  successor is not consistent with the choice from $(\hat{v},1)$ we
  update the memory according to the choice from $(\hat{v},0)$.

  Consider a play in $M$ consistent with this strategy. Clearly, a
  version of this play appears also in $G$.
  As both the game and the MDP agree on $W_s$ and the strategy in $G$
  is winning it follows that $W_s$ is satisfied in $M$ as well.
  Consider now the winning condition $W_{as}$.
  If the play visits only finitely many configurations in $V_p$ then
  it is identical to the play in $G$ and satisfies also $W_{as}$.
  If the play visits infinitely many configurations in $V_p\cap B$
  then it is clearly winning according to $W_{as}$.
  Otherwise, the play visits infinitely many configurations in $V_p$
  but only finitely many in $V_p\cap B$.
  The play cannot stay forever in a non-bottom SCC of $G$ (combined
  with the strategy of \pzero).
  Indeed, there is a non-zero probability for the probabilistic
  choices to agree with the strategy of \pzero and leave this
  non-bottom SCC.
  Similarly, when reaching a bottom SCC, as all configurations of the
  SCC will be visited with probability 1 and from the strategy being
  winning in $G'$ we conclude that there is a configuration in $B$
  appearing in this bottom SCC and it will be visited infinitely
  often. 
\end{proof}

\shortv{
  \noindent
  We include the proof of Lemma~\ref{lemma:buchi and parity to
    parity}:

  \proofoflemmabuchiandparitytoparity
}

\subsection{Proofs from Section~\ref{section:sas games}}

\shortv{
  \noindent
  We include the proof of Theorem~\ref{theorem:sas games determinacy}:

  \proofoftheoremsasgamesdeterminacy
}

\shortv{
  \noindent
  We include the proof of Theorem~\ref{theorem:sas games memoryless
    player one}:

  \proofoftheoremsasgamesmemorylessplayerone
}

\shortv{
  \noindent
  We include the proof of Lemma~\ref{lemma:good ranking induces as
    ranking}:
  \proofoflemmagoodrankinginduces
}

\shortv{
  \noindent
  We include the proof of Lemma~\ref{lemma:good as ranking induces}:
  \proofoflemmagoodasrankinginduces
}

%% file: figures/gadgetmdp.tex
\game{
\po{1}{}{\overline{v}}
\labeln{1}{right=0.1mm of 1,xshift=-7mm,yshift=5mm}{$v\in B$}
\pz{2}{below of=1}{(\tilde{v} ,0)}
%\pz{3}{right of=2,xshift=7mm}{(\tilde{v} ,2)}
\po{6}{below of=2}{(\hat{v} ,0)}
%\pz{7}{right of=6}{(\hat{v} ,1)}
%\po{8}{right of=7}{(\hat{v} ,2)}

\edgestosucc{6}
%\edgestosucc{7}
%\edgestosucc{8}

% This is the draft for edges to succ macro:
%\node (ls1) [below of=6,yshift=2mm] {$succ(v)$};
%\node (ls1r) [right=0.1mm of ls1,yshift=-2mm] {};
%\node (ls1l) [left=0.1mm of ls1,yshift=-2mm] {};
%\edge{6}{ls1r}
%\edge{6}{ls1l}

\edge{1}{2}
%\edge{1}{3}
\edge{2}{6}
%\edge{3}{7}
%\edge{3}{8}

\po{9}{right of=1,xshift=30mm}{\overline{v}}
\labeln{9}{right=0.1mm of 9,xshift=-7mm,yshift=5mm}{$v\notin B$}
\pz{10}{below of=9,xshift=-10mm}{(\tilde{v} ,0)}
\pz{11}{right of=10,xshift=7mm}{(\tilde{v} ,2)}
\po{12}{below of=10}{(\hat{v} ,0)}
\pz{13}{below of=11}{(\hat{v} ,1)}

\edgestosucc{12}
\edgestosucc{13}

% This is the draft for edges to succ macro:
%\node (ls1) [below of=6,yshift=2mm] {$succ(v)$};
%\node (ls1r) [right=0.1mm of ls1,yshift=-2mm] {};
%\node (ls1l) [left=0.1mm of ls1,yshift=-2mm] {};
%\edge{6}{ls1r}
%\edge{6}{ls1l}

\edge{9}{10}
\edge{9}{11}
\edge{10}{12}
\edge{11}{13}
}